\documentclass[pdftex,twocolumn,epjc3]{svjour3}          

\RequirePackage[T1]{fontenc}

\smartqed  

\RequirePackage{graphicx}
\RequirePackage{mathptmx}  
\RequirePackage{flushend}
\RequirePackage[numbers,sort&compress]{natbib}
\RequirePackage[colorlinks,citecolor=blue,urlcolor=blue,linkcolor=blue]{hyperref}

\journalname{Eur. Phys. J. B}
\graphicspath{{Plots/}}

\newcommand{\brac}[1]{\left( #1 \right)}  
\newcommand{\avg}[1]{\left\langle #1 \right\rangle}  
\newcommand{\abs}[1]{\left\vert #1 \right\vert}  

\begin{document}

\title{Phase transition in the bipartite \texorpdfstring{$z$}{z}-matching}

\author{Till Kahlke\thanksref{e1,addr1} 
\and
Martin Fr\"anzle\thanksref{e2,addr2}
\and
Alexander K. Hartmann\thanksref{e3,addr1}
}%
\thankstext{e1}{till.kahlke@uol.de}
\thankstext{e2}{martin.fraenzle@informatik.uni-oldenburg.de}
\thankstext{e3}{a.hartmann@uni-oldenburg.de}

\institute{Institut f\"ur Physik, Carl von Ossietzky Universit\"at 
Oldenburg, 26111 Oldenburg, Germany \label{addr1}
\and
  Department f\"ur Informatik,  Carl von Ossietzky Universit\"at Oldenburg, 
26111 Oldenburg, Germany \label{addr2}
}

\date{Received: date / Accepted: date}

\abstractdc{
  We study numerically the maximum $z$-matching problems on ensembles
  of bipartite random graphs. The $z$-matching problems describes
  the matching between two types of nodes, \emph{users} and \emph{servers}, 
where each server may   serve up to $z$ users at the same time.
  By using a mapping to standard maximum-cardinality
  matching, and because for the
  latter there exists a polynomial-time 
exact algorithm, we can study large system
  sizes of up to $10^6$ nodes. We measure the capacity and the energy
  of the resulting optimum matchings. First, we confirm previous analytical
  results for bipartite regular graphs. Next, we study the finite-size
  behaviour of the matching capacity and find the same scaling behaviour
  as before for standard matching, which indicates the universality of
  the problem. Finally, we investigate for bipartite Erd\H{o}s-R\'enyi
  random graphs the saturability as a function of the average degree,
  i.e., whether the network allows as many customers as possible to be served,
  i.e. exploiting the servers in an optimal way.
  We find phase transitions between unsaturable and
  saturable phases. These coincide with a strong change of the running time
  of the exact matching algorithm, as well with the point where
  a minimum-degree heuristic algorithm starts to fail.
}

\maketitle

\section{Introduction}

Phase transitions in combinatorial optimisation or
in con\-straint-satisfaction
problems
\cite{martin2001,phase-transitions2005,moore2011,mezard2009} have been
an active area of research
at the interface of statistical mechanics and computer science since
more than two decades. Usually,
non-deter\-ministi\-cally polynomial (NP) complete \cite{garey1979}
or NP-hard  problems are studied, i.e.,
problems for which so far no algorithm is known which runs
in the worst case in polynomial time as a function of the system size. Thus,
so far only worst-case exponential-time algorithms are available,
i.e., the problems are \emph{hard}.
One the other hand,
problems running in polynomial (P) time are often termed \emph{easy}.
But since
it is not known, i.e., proven, whether P forms a proper subclass of NP,
or whether maybe P=NP, one has been interested since almost
the beginning of computer science in finding out what makes
a problem hard. Among many other approaches, also numerical
experiments and statistical-mecha\-nics calculations have been performed.
For this purpose, ensembles of random problems have been
considered, which are hard in the worst case but 
for some regions in parameter space
typically require only a polynomial running time, meaning they are typically 
easy there.
Here in particular phase transitions \cite{monasson1999,martin2001},
e.g., with respect
to the solvability
have been observed  when varying suitable ensemble parameters. These
phase transitions often coincide with changes of the typical complexity
from easy to hard. Thus, the structure of problems from such ensembles,
in particular near phase transitions, may teach us about the source
of computational hardness. Such phase transitions have been studied,
e.g., for constraint satisfaction problems like
satisfiability (SAT) \cite{kirkpatrick1994,cocco2001} or
colouring \cite{mulet2002}. Also optimisation problems like the
travelling salesperson
 \cite{mezard1986,gent1996tsp,tsp_lp2016,tsp_rsb2019},
vertex cover  \cite{cover2000,cover-tcs2001,cover-time2001,cover-long2001}
or  number partitioning \cite{mertens1998,mertens2006} have been investigated.
Beyond delivering insight into the structure of problems,
this reserach performed at the interface of physics and computer science
has also led to algorithmic advances like the
development of efficient message-passing algorithms as
Belief Propagation or Survey Propagation \cite{mezard2002}.

Nevertheless, not only hard optimisation or constraint-satisfaction
problems may exhibit changes
of problem space structure and corresponding changes of the computational
complexity. Also ensembles of polynomially-solvable problems like
shortest paths, maximum flows or graph matching
may be of interest and show corresponding phase transitions.
Usually,  in physics such algorithms
are used to investigate models like random magnets
\cite{opt-phys2001,opt-phys2004}. Here, we want to perform a fundamental
study of such a phase transitions for a generalisation of the
graph matching problem. As we will see here, we observe 
changes of the algorithmic behaviour
related to this phase
transition. This shows
that such a coincidence of the change of a suitably defined
solvability and of algorithmic
complexity exists  also for a polynomially-solvable problem and can therefore
easier be studied numerically.

For a given graph, 
the maximum-cardinality matching problem, also just called matching,
considers subsets of edges, such that
each node is incident to at most one edge in the subset
and such that the cardinality of matching is maximum.
This problem
is  widely studied in computer science usually
from the algorithmic point of view \cite{jungnickel2010} with the aim
to find efficient algorithms.
But also in the field of statistical mechanics it has already played its
role, as it
was among the first optimisation
problems studied, and therefore it  has inspired the field a lot.
First, the model was solved analytically
  using a replica-symmetric
  approach \cite{mezard1985,parisi2002}
on bipartite random graphs with random edge weights, i.e.,
for the maximum-weight matching instead of the maximum-cardinality matching.
 This suitability of a replica-symmetric calculation 
means that the  thermodynamic behaviour
  of matching for this ensemble is not very complex, similar to a ferromagnet.
  Later, the solution was extended to arbitrary graphs
  and the finite-size behaviour
  of the matching capacity, i.e., the sum of the weights
  of the edges in the matching, was obtained \cite{mezard1987matching}.
  Also studied were
  Euclidean variants \cite{mezard1988,houdayer1998,caracciolo2014}.
 The case
  where more than two nodes are connected per matching element was
  also considered
  with a statistical mechanics approach \cite{martin2005}.
  Furthermore, so called \emph{dimer coverings}, i.e., perfect
  matchings involving all nodes, on
   $d$-dimensional lattices without edge weights
  were studied \cite{kenyon2009}.
   Since no energy is involved, the number
   of matchings, i.e., the entropy, was mainly studied
   \cite{kasteleyn1961,fisher1961,tempereley1961}.
   Such studies of entropies of dimer coverings
   were extended also to include
   energy for the edges  \cite{fisher1966}, or
   to  mixtures
   of dimers and single atoms  \cite{heilmann1972}.
   Also, entropies of the matchings \cite{karp1981}
   or dimer coverings \cite{zdeborova2006}
   were considered for various more general random graph
   structures.

In this work, we study a phase transition of the
satisfiable-unsatisfiable type for
the \emph{$z$-matching problem}, which is a generalisation
of the standard matching. The model describes
a set of $N$ \emph{users} and a set of $S$
\emph{servers}, possible user-server
connections are described by a bipartite graph.
Each user shall be served by one server, while 
each server may serve up to $z$ users at
the same time. The system is characterised by its capacity, i.e.,
the number of users which can be served simultaneously .
An example for the application of $z$-matching are wireless
communication networks \cite{gu2015,han2017}.
Our study is motivated by a previous work
of Krea\v{c}\i\'c and Bianconi \cite{Kreacic_2019}, who have studied,
to our knowledge for the first time in statistical mechanics,
the $z$-matching
problem analytically with the approximate cavity approach and numerically
with a message-passing
algorithm. They have obtained the capacity of the system
for two ensembles, namely for
fixed degree and Poissonian bipartite
graphs. They showed that for both cases, parameter combinations exists,
where the capacity converges to its maximum possible value, i.e.,
a \emph{saturable} phase, when increasing the average node degree.

Here, we expand on this work by using an exact numerical matching algorithm.
Since this algorithm allows for calculating exact optimum
matchings in polynomial time, we are
able to solve exactly very large graphs of up to $N=500000$ users.
To start, we confirm with our exact numerical
approach the previously obtained analytical results from
Ref.~\cite{Kreacic_2019}. Also, we find the same finite-size scaling
behaviour of the capacity as for standard matching
\cite{mezard1987matching}. In the main part of our work,
for the case of the Poissonian random graphs with average user degree $k$,
we investigate the model with respect to the phase transition between
saturability and unsaturability
for some typical  parameter combinations
of $z$ and the ratio $N/S$. We determine with high precision the
phase-transition point $k_c$ using finite-size scaling techniques.
In addition, we
obtain the critical exponent $\nu$ characterising the phase transition.
By analysing the run-time of the exact algorithm and furthermore studying
an approximation algorithm, we are able to show that the phase
transition coincides with remarkable changes of the algorithmic
performance.

The remainder of the paper is organised as follows: Next, we define
the $z$-matching problem and the measurable quantities we have evaluated,
together with the ensemble of random graphs we have studied. In the third
section, we present the methods we have applied. In the main section,
we present our results and finally, we summarise our work and
outline further research directions.

\section{Model}

\subsection{\texorpdfstring{$z$}{z}-matching}

We consider bipartite graphs $G = (V, E)$ with vertices $i\in V = A \cup B$
consisting of $N = |A|$ users and $S = |B|$
servers.
We denote the ratio of number of users and servers by
\begin{equation} \label{eq:def:eta}
  \eta = N/S\,.
  \end{equation}
Since the graph is bipartite, each edge connects a user and a server. This means that the average degree $k$ of the users  and the average degree $q$ of the
severs  are related by
\begin{equation}
  N k = S q.
  \label{eq:Nk=Sq}
\end{equation}
When inserting \autoref{eq:def:eta}, it follows that $q = \eta k$.

A \emph{$z$-matching} $M_z$ is a subset of $E$ in with each user is adjacent to at most one server and each server is adjacent to at most $z$ users. For each edge in $M_z$ we say that the adjacent user and server are
\emph{matched}.
Thus, one server can be matched to at most $z$ users.
The \emph{capacity} $C$ of the matching 
is defined as the number $\abs{M_z}$ of edges in the
matching, i.e., its cardinality.
In accordance with previous work \cite{Kreacic_2019},
the \emph{energy} $H$ is defined as
\begin{equation}
  H = zS + N - 2C .
  \label{eq:H}
\end{equation}
Hence, $H$ is equal to the number of unmatched users plus the number of
users that all servers can still host.
A low value of the energy
means that the given resources are well used, such that many, or all users
are served and at the same time not too many, possibly no, 
servers exhibit unused serving capacities.

For the maximum $z$-matching problem one wants find a
$z$-matching $M_z$ which maximises the capacity $C$. This will
usually depend on the edges in the graph.
For any graph, the capacity 
 $C$ is bounded by a  theoretical capacity which
is obtained, when all users are
matched to servers or when all servers are matched to $z$ users.
This leads to the maximum
theoretical capacity
\begin{equation}
  C_\mathrm{max} = \min\brac{N, zS} = \min\{1,z/\eta\}N\,.
  \label{eq:C_max}
\end{equation}
Note that $C_\mathrm{max}$ does not at all depend on the edges, i.e.,
the actual topology of the graphs.
But
it depends on the given graph $G$ whether this maximum can actually 
be reached. Naturally, the more edges exist, the more likely
it is that  a given graph  reaches $C_\mathrm{max}$.
For numerical reasons, we do not require that $C_\mathrm{max}$
is 100\% reached. Instead,
we call a graph \emph{saturated} if the actual capacity reaches
$\gamma C_{\max}$ with $\gamma \le 1$ being a suitable threshold.
Correspondingly, when studying an ensemble of random graphs,
the saturation probability  is defined as
\begin{equation}
  p_\mathrm{sat} = \mathrm{Prob}\brac{C \geq \gamma \ C_\mathrm{max}}.
  \label{eq:p_sat}
\end{equation}
We use $p_\mathrm{sat}$ as  an order parameter. If $p_\mathrm{sat}$ is
close to zero, few graphs reach the capacity $\gamma C_{\max}$.
We call this the
unsaturated phase. But if $p_\mathrm{sat}$ is close to one, almost all
graphs have a high capacity. This describes the saturated phase. As our results
will show below, we are indeed able to observe phase transitions between
these two phases.

In principle one could use $\gamma=1$, but for large random graphs,
for most ensembles, it can be anticipated
that it is exponentially unlikely that all demands
can be matched exactly to all resources. We confirmed this
in our numerical
experiments.  Thus, we used a value of $\gamma$ close to 1, i.e.,
$\gamma=0.9$, hence we call a network saturated if there is an almost
complete balance between demand and resources.
We also verified by tests that our results did not change
significantly when we used other values like $\gamma=0.95$ or $\gamma=0.85$.

Note that
the variance of $p_\mathrm{sat}$
is given simply by
$\sigma^2_{p_\mathrm{sat}}$ $=$ $p_\mathrm{sat} \brac{1 - p_\mathrm{sat}}$.
We used it to calculate our error bars and to obtain more conventiently
the positions of the phase transitions, see \autoref{sec:pue}.

\subsection{Random graphs}
We consider the same  two different networks ensembles as previously studied
\cite{Kreacic_2019}.
The first networks ensemble consist of ($k,q$)-regular graphs were all users have the same fixed integer degree $k$ and all servers have integer degree $q$, satisfying \autoref{eq:Nk=Sq}.
To generate such random graphs for the numerical studies, the configuration model \cite{Bollobas_1980, Newman_2001} can be used.
In our case for ($k,q$)-regular graphs we did the following:
$k$ \emph{stubs} are assigned to each user and $q$ stubs to each server. Then
iteratively one stub from a user and one stub from a server are drawn randomly (with uniform distribution from the list of free stubs), respectively.
If there is so far no edge between the two nodes where the stubs belong to, the edge connecting them
is created and the used stubs get removed. If this edge already exists, two new stubs are drawn. This procedure is repeated until no stub is left.
According to \cite{Klein-Hennig_2012}, redrawing two stubs if the edge already exists create a bias in the generation of the random graphs.
This bias can be removed,  after the initial graph construction
has finished, by repeatedly swapping the edges of the
generated graph \cite{Viger_2005}. For a swap, two edges are chosen
randomly. Then the users, but not the servers to with the edges are
connected are swapped. Repeating this `shuffle` enough times will
lead to unbiased graphs.
We tested the influences of edge swaps and in our case, they had no
measurable effect on the final data. This might be because we consider
bipartite graphs.
Thus, to save computation time, we   did not perform  such swaps.

The second network  ensemble we considered, which is in the center of our 
study, consist of bipartite
Erd\H{o}s-R\'{e}nyi graphs \cite{Edroes_1959}. This means, each possible
edge between a user and a server is drawn with a probability $p$,
with $p=k/S$ for any desired average number $k$ of neigbouring servers
for the users. Thus,
the degree distributions of users and servers  are Poissonian, respectively.

\section{Methods}

\begin{figure}
\includegraphics{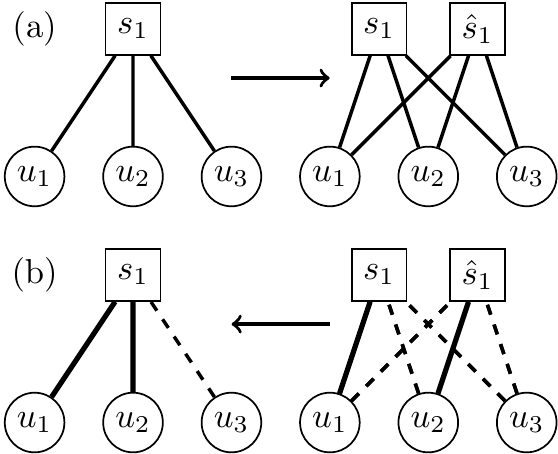}
\caption{\label{fig:clone-nodes} (a) Cloning the server $s_1$ 
for the case $z=2$ to create $\hat{s}_1$ having the same neighbours 
as $s_1$. (b) A matching on the graph with cloned servers corresponds 
to a $z$-matching on the original graph. Edges in the matching 
are marked bold, non-matched edges are marked dashed.}
\end{figure}

To solve the maximum $z$-matching problem, which we call also just
$z$-matching in the following,
numerically on a given graph, we map the problem to the original
1-to-1-matching problem on bipartite graphs.
To achieve this, each server node is cloned $z-1$ times. This means that
$z-1$  new nodes will be inserted in the graph for every server
node. Then, for each user in the neighbourhood of the server, an edge
to each of the new $z-1$ nodes is created. Thus,
in the graph with cloned nodes, each
clone has the same neighbours as the original server node.
Next, a matching is calculated for the graph with the cloned nodes.
Here each node is matched at most once.
This means, with respect to the original graph, each user
node will be matched at most once and each server node will be matched
at most $z$ times.
Therefore, a standard 1-to-1-matching on the graph with cloned nodes
corresponds to a $z$-matching on the original graph.
This procedure is shown in \autoref{fig:clone-nodes}.

To find a maximum matching we used Edmond's Blossom Shrinking algorithm \cite{Cook_1998}, implemented in the LEMON-library \cite{LEMON}.

\section{Results}

We have performed simulations \cite{practical_guide2015}
by using exact numerical matching calculations
for the two graph ensembles, for various values of the parameters,
for various graph sizes of up to $N=500000$ user nodes and
up to $S=250000$ server nodes. We performed for all results
an average over up to several thousands of
different graph realisations. Details
are stated below. We first compare our numerical results with the
previously obtained analytical results \cite{Kreacic_2019}. Next,
we investigate the finite-size scaling behaviour of the capacity
and compare with scaling form previously found \cite{mezard1987matching}
for standard matchings.
In the main part, we show the results concerning the saturable-unsaturable
phase transition and compare with the algorithmic behavior.

\subsection{Comparison with previous results}

\begin{figure}
  \includegraphics{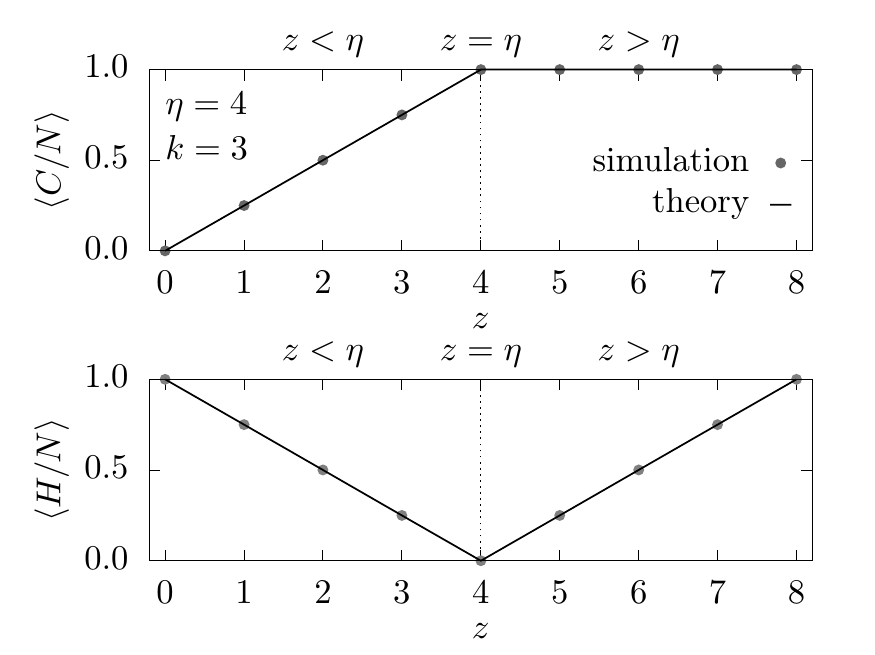}%
  \caption{\label{fig:Delta_C_and_H_over_z} Capacity density $C/N$ and
    energy density $H/N$ of the $z$-matching as a function of $z$ for
    ($k,q$)-regular graphs with $N = 40000$ user nodes.
    The circles denote the results of the simulation and the line
    represents the analytical predictions from \cite{Kreacic_2019}.
    The lines are actually guide to
    the eyes which connect the predictions which are available
    only for integer values of $z$ as well.
    The dashed vertical line marks $z = \eta$. There are no error bars
    because the results are always identical.}
\end{figure}

We studied the ($k,q$)-regular graphs, i.e, for integer values of $k$.
We considered a rather
large graph size $N=40000$ and $\eta=4$, thus $S=10000$. We have considered
$k=3$, thus $q=\eta k = 12$.
In \autoref{fig:Delta_C_and_H_over_z} the average capacity per user node $\langle C/N \rangle$
and the average energy per user node $\langle H/N \rangle$ are shown  as a function of $z$.
The results are averaged over 100 realisations for each value of $z$.
But due to the simple structure of the graphs, although being random,
there are no statistical fluctuation on the results.

One can observe three cases,  dependent on the value  of  $z$:
\begin{itemize}
\item   For $z < \eta$, or equivalently $z k < q$, the capacity is
  always $C = z S = C_\mathrm{max}$ and the energy is $H = N - z S > 0$.
  All servers are matched to $z$ users, but there are more users than
  all servers together can handle.
\item At $z = \eta$, i.e., $z k = q$, the capacity is
  $C = N = C_\mathrm{max}$ and the energy is $H = 0$. This is an optimal
  situation since all users are matched to servers and no server has
  unused resources.
\item
  For $z > \eta$, i.e., $z k > q$, the capacity is
  $C = N = C_\mathrm{max}$ and the energy is $H = - N + z S > 0$.
  All users are matched, but there are more servers than needed.
\end{itemize}
  Thus, in all networks, the full capacity is reached, but there is only
  one point of optimal balance between user demands and provided resources,
  where the energy is zero. It is interesting, that such a point
  of balance is possible. Note that here with $k=3$, the graphs
  exhibit a rather large number of edges
  much larger than the number of edges in the matching. On the other
  hand for the smallest meaningful degree $k=1$,
  at $z=\eta$ all users will be connected
  to exactly one server, and, because of $q=\eta k=z$, each server
  will be connected to $z$ distinct nodes. Thus, trivally the full edge
  set is a $z$-matching and again all user demands are satisfied.
  Anyway,
  these results confirm those recently obtained analytically
  \cite{Kreacic_2019}
  for this network ensemble.


\begin{figure}
  \includegraphics{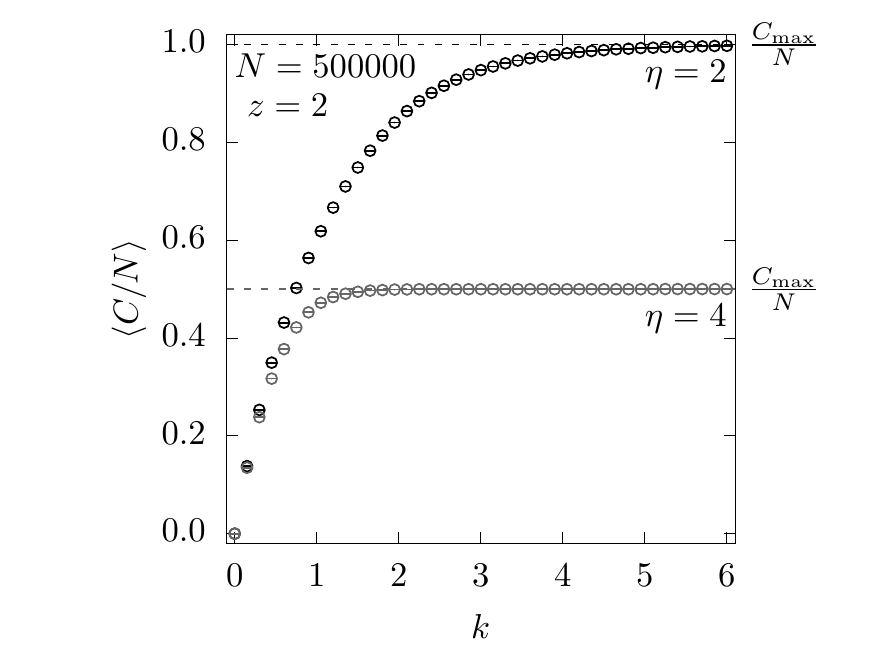}%
  \caption{\label{fig:Poisson_c_over_k} Average capacity density $\avg{C/N}$
    as a function of the mean user degree $k$ for bipartite Erd\H{o}s-R\'{e}nyi graphs, $N=500000$, $z=2$, for
    two cases $\eta=2$ and
    $\eta=4$. The error bars are smaller than the point size.
  }
\end{figure}

  More variations in the results are obtained for the other network
  ensemble we have studied, the bipartite
  Erd\H{o}s-R\'{e}nyi graphs, since for this ensemble the nodes
  exhibit fluctuations of the degrees.
  In \autoref{fig:Poisson_c_over_k} the average capacity $\langle C/N \rangle$
  per user node is shown as a function of the average user degree $k$
  for server capacity $z=2$ and $N=500000$ users.
  The results are averaged over 500 realisations for each value of $k$.
  Two cases for the user to server ratio are considered, $\eta=2$
  and $\eta=4$. These values correspond to average server degrees
  $q=2k$ and $q=4k$, respectively and server numbers
  $S=N/\eta= 250000$, and $S=125000$, respectively.
  When approaching large degrees $k$, in
  both cases the limiting capacities are reached which are
  according to \autoref{eq:C_max}
  $C_{\max}/N=1$ for $\eta=2$ and $C_{\max}/N=0.5$ for $\eta=4$.
  Our results from using the exact algorithm
  agree well with the previous results \cite{Kreacic_2019}
  obtained by a message-passing algorithm.
  Note that we obtained data for all values of $q$ down to zero,
  while in the previous work only the range $q >z$, i.e. $k >z/\eta$
  was considered. Anyway, this success of the message-passing
  approach in the previous work is interesting, because
  it is known that for other problems,
  like the NP-hard vertex-cover problem,
  message-passing fails in the range of high degrees
  because for such models there exist replica-symmetry breaking
   \cite{cover2000,cover-tcs2001,cover-long2001}.
   The reason for the success with respect to $z$-matching
   could be that in the range of large values
  of $k$, the problem is easy to solve since there are enough options
  for each user and each server, which do not block each other too much.
This corresponds to a simple, i.e, ``dense'' organization of the solution
space and makes a quick convergence of the message-passing iterations 
possible.
  On the other hand for NP-hard problems, each assignment
of a problem variable  has typically a strong
  impact on the availability of suitable assignments for other variables.

\subsection{Finite-size behaviour}
Since there are basically no finite-size effects for the
($k,q$)-regular graphs, we studied the finite-size scaling behaviour for
bipartite Erd\H{o}s-R\'{e}nyi graphs.
For $N \leq 1000$ we averaged over 5000 different realisations and for
larger values of $N$, a number of 500 realisations turned out to be sufficient.

\begin{figure}
  \includegraphics{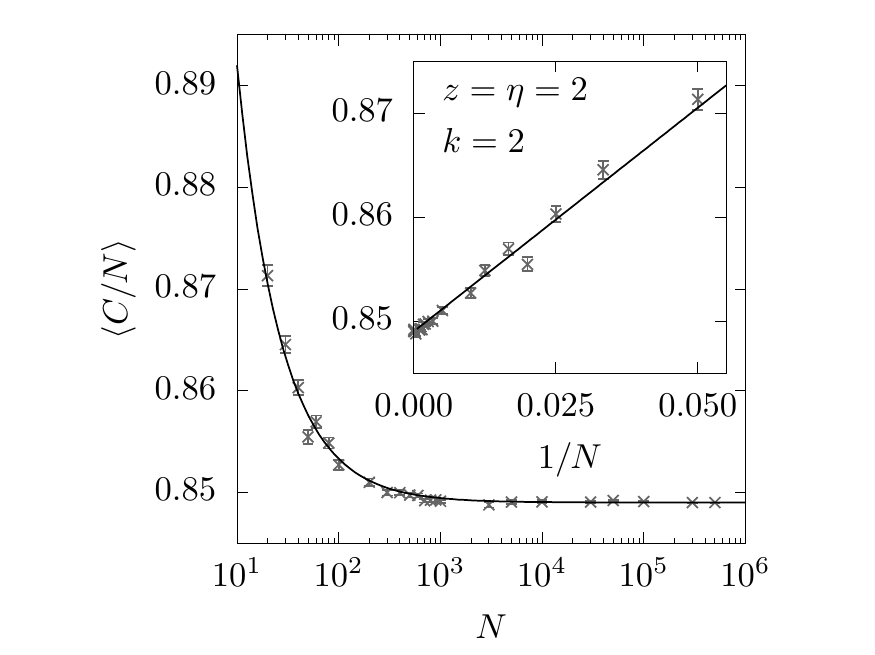}%
  \caption{\label{fig:Poisson_c_over_N} The average capacity
    density $\avg{C/N}$ as a function of the number of users $N$
    for bipartite Erd\H{o}s-R\'{e}nyi graphs.
    The line represents the fit according to \autoref{eq:fit_c_N}.
    Inset shows $\avg{C/N}$ as function of $1/N$ for the same data to confirm the
    $1/N$-behaviour.}
\end{figure}

In \autoref{fig:Poisson_c_over_N}, the average capacity density
$\avg{C/N}$ is shown as a function of the system size for the case $k=2$,
$\eta=2$ and $z=2$. Motivated by the results \cite{mezard1987matching}
for standard matching, we fitted the data to the function
\begin{equation}
  \avg{C/N} = c_{\infty} + \alpha_1 \frac{1}{N}\,
  \label{eq:fit_c_N}
\end{equation}
and found a good agreement.
The resulting parameters, also for the other cases we have studied,
are shown in \autoref{tab:fitpara_c_over_N}.
For $\eta = 4$, $z = 2$ and $k = 3$ the fit does not converge since most
graphs reached $C_\mathrm{max}$ and therefore a finite-size dependence is
hardly to observe. For all other cases, we observe a good agreement with
the $1/N$ scaling. Only the prefactor of this term seems to be non-universal.
Also it should be noted that not only the finite-size capacities but
also the limiting values of the capacity density are usually well below
$C_{\max}/N$,
due to typically too small average degrees.
This corresponds also to the behaviour seen in \autoref{fig:Poisson_c_over_k}.

\begin{table}
  \caption{\label{tab:fitpara_c_over_N} Maximum capacity $C_{\max}$ and
    fit parameters for the scaling behaviour \autoref{eq:fit_c_N}
    for different values of the parameters $\eta$, $z$ and $k$.}
  \begin{tabular}{c c c l l l}
        {$\eta$}  & {$z$}  & {$k$}   & $C_{\max}/N$ & {$c_\infty $} & {$\alpha_1$} \\
        \hline
        2 & 2 & 1 & 1 & 0.60089(2)  & 0.32(1) \\
        ~ & ~ & 2 & 1 & 0.84902(2)  & 0.43(1) \\
        ~ & ~ & 3 & 1 & 0.947900(9)  & 0.388(6) \\
        4 & 2 & 1 & 0.5 & 0.466631(9)  & 0.118(8) \\
        ~ & ~ & 2 & 0.5 & 0.499142(1)  & 0.0141(6) \\
        ~ & ~ & 3 & 0.5 & {--} & {--}  \\
        4 & 4 & 1 & 1 & 0.62561(2)  & 0.70(2) \\
        ~ & ~ & 2 & 1 & 0.86292(2)  & 1.09(1) \\
        ~ & ~ & 3 & 1& 0.950115(9)  & 0.794(6) \\
  \end{tabular}
\end{table}

\subsection{Phase transition \label{sec:pue}}

Finally, we study the saturation probability $p_\mathrm{sat}$.
Due to the simple structure of ($k,q$)-regular graphs,
we only focus on the Erd\H{o}s-R\'{e}nyi graphs and investigate them
while varying the average degrees $k$ and $q=\eta k$.
All results are obtained over 500 difference realisations
for each value of $k$ and each system size $N,S=N/\eta$.

\begin{figure}
  \includegraphics{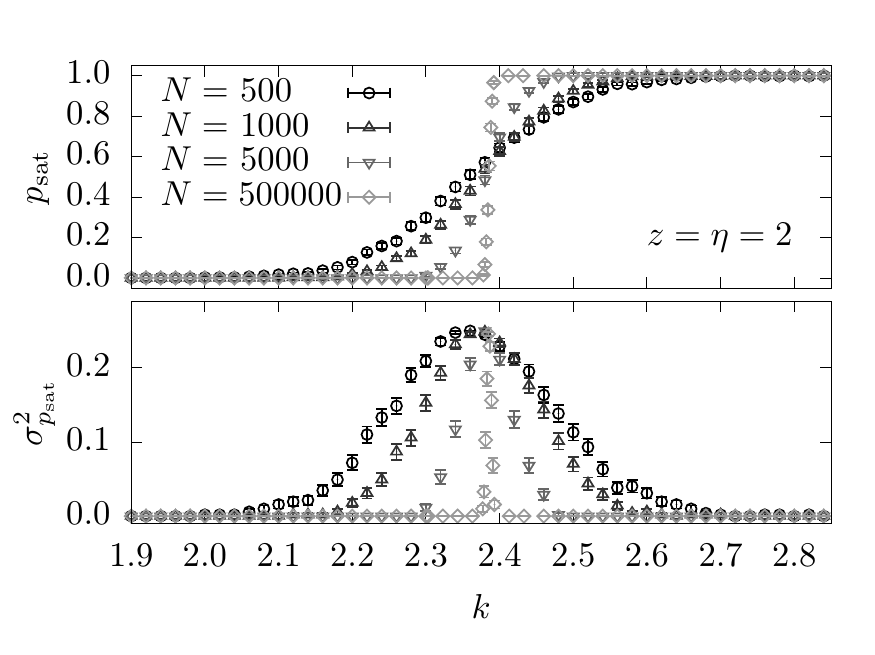}%
  \caption{\label{fig:Poisson_p_sat_over_k} Saturation probability
    $p_\mathrm{sat}$ (top) and its variance $\sigma^2_{p_\mathrm{sat}}$
    (bottom) as a function of $k$ for different number $N$ of users.
    For the largest number of users there is an almost step-wise
    transition from $p_\mathrm{sat} = 0$ to $p_\mathrm{sat} = 1$ at
    some value of $k = k_c$.}
\end{figure}

\autoref{fig:Poisson_p_sat_over_k} shows $p_\mathrm{sat}(k)$
for different number $N$ of users.
$p_\mathrm{sat}$ increases from 0 to 1 with growing degree $k$,
such that the curves become steeper when increasing $N$.
This is an indication for  a transition from an unsaturated phase to a
saturated phase. Note that this transition becomes almost step-wise for a
really large number of users. We are able to observe the phase transition
in such a clear way, because we could study huge system sizes due to
the polynomial nature of the problem. This is in contrast to
previously studied phase transitions for NP-hard optimisation problems,
where only exponential-time exact algorithms are known and therefore only
rather small system sizes could be studied exactly.

\begin{figure}
  \includegraphics{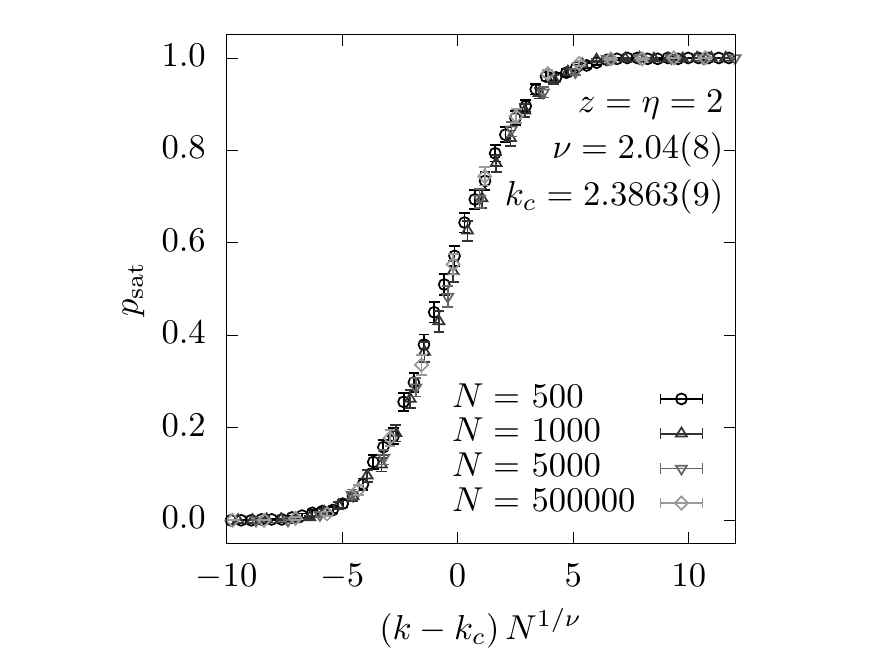}%
  \caption{\label{fig:Poisson_fss} Finite-size scaling for the saturation
    probability. Within the statistical fluctuations, the values for
    different $N$ overlap. For visualisation only four system sizes $N$
    are shown, but the results were obtained including all measured sizes $N$.
  }
\end{figure}

\begin{sloppypar}
To study the observed phase transition in more detail,
we use finite-size scaling analysis \cite{Cardy_2015_fss}.
Hence, we assume that for continuous transitions
the saturation probability follows the standard
 finite-size scaling relation
\begin{equation}
p_\mathrm{sat}(k,N)=\tilde p_\mathrm{sat}((k-k_c)N^{1/\nu})\,,
  \end{equation}
with infinite-size critical point $k_c$ and $\nu$ being the exponent
describing the divergence of the correlation length.
We obtained the best-fitting scaling parameters using the
tool {\tt autoScale.py} \cite{Melchert_2009}.
The results are obtained over 9 different system sizes $N$,
ranging from 500 to 500000.
\autoref{fig:Poisson_fss} shows the resulting data collapse for
$\eta =2$, $z = 2$. Apperently the collapse works very good for this case.
The best found values for $k_c$ and $\nu$ for all studied cases can
be found in \autoref{tab:fss}.
The quality $\mathcal{S}$ is the average deviation of the data points
from the collapse curve, measured in terms of error bars \cite{gauss_2d}.
Since $\mathcal{S}$ is close to one for all cases we have considered,
the quality of the data collapse is always very good.
\end{sloppypar}

\begin{table}
  \caption{\label{tab:fss} Results for the finite-size scaling parameters
    $k_c$, $\nu$
    and the quality $\mathcal{S}$ of the fit, for the considered cases
    of $\eta$ and $z$.}
  \begin{tabular}{c c c c c}
      {$\eta$} & {$z$} & {$k_c$} & {$\nu$} & {$\mathcal{S}$}\\
      \hline
      2 & 2 & 2.3863(9) & 2.04(8) & 0.90 \\
      4 & 2 & 0.8852(4) & 2.0(1)  & 0.84 \\
      4 & 4 & 2.312(1)  & 2.0(1)  & 1.48 \\
    \end{tabular}
\end{table}
The statistical errors of the scaling parameters are determined as how much
a parameter has to be changed to increase the quality $\mathcal{S}$ 
of the fit by one to $\mathcal{S}+1$.
We also systematically changed the intervals over which the data collapse
is performed. However, these differences turned out to be
smaller than the statistical errors, so we state only those.
Interestingly, within error bars, the value for $\nu$ is compatible
with a value of $\nu=2$ in all
studied cases. This indicates that the behaviour of the
saturable-unsaturable phase transition is universal with respect to
network parameters.

\begin{figure}
  \includegraphics{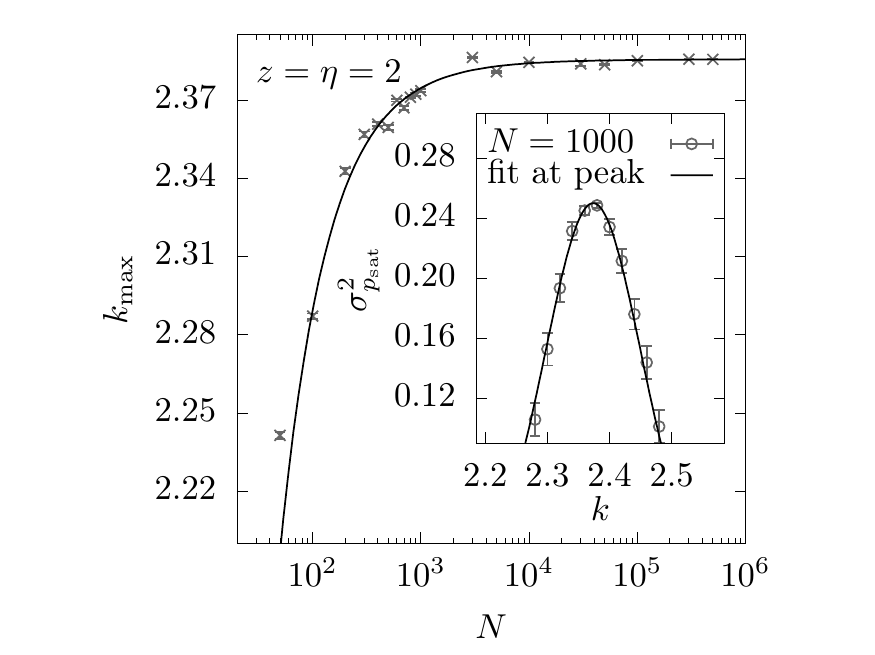}%
  \caption{\label{fig:Poisson_k_max_over_N} Position of maximal variance $k_\mathrm{max}$ as a function of  the number of users $N$. The line represents the fit. The inset shows an example for the Gaussian fit around the peaks of the
    variance $\sigma^2_{p_\mathrm{sat}}$ of the saturation probability.
    Note that
    the peak position shifts to the right for larger values of $N$.}
\end{figure}

Note that the phase transition can also be studied and analysed
by obtaining the variance $\sigma^2_{p_\mathrm{sat}}$ as a function
of $k$. Although  $\sigma^2_{p_\mathrm{sat}}$ does not contain any
additional information, it is easy to analyze, because it
peaks at the apparent  transition for each system size, which can be seen in
the bottom of \autoref{fig:Poisson_p_sat_over_k}.
To confirm the results obtained from $p_\mathrm{sat}$,
we also analysed the phase transition
by a finite-size analysis of the variance.
In the thermodynamic limit, the variance should be maximal
at the critical point. On finite systems, the position of maximal variance,
denoted as $k_\mathrm{max}$, will approach  $k_c$ as the system size grows.
To determined $k_\mathrm{max}$ more accurately than given by
the resolution of the considered values of $k$, we performed Gaussian fits
to $\sigma^2_{p_\mathrm{sat}}(k)$ in small intervals near
the maxima. An example is shown in the inset of
\autoref{fig:Poisson_k_max_over_N}.
To extrapolate $k_\mathrm{max}$ the fit
\begin{equation}
  \label{eq:fit_k_max}
  k_\mathrm{max}(N) = k_c + \beta_1 ~ N^{-1/\nu} \brac{1 + \beta_2 ~ N^{-1/\nu}}
\end{equation}
is used. Note that we had to use here a correction term to the
scaling behaviour, taking care of the very small system sizes.
But we did not need to add a correction exponent to achieve a good fit
and used exponent $2/\nu$ instead.
Also, based on the above results of the finite-size scaling, we
fixed $\nu = 2$.
\autoref{fig:Poisson_k_max_over_N} shows the fit for the case
$\eta = z= 2$.
The results of the obtained fit parameter for all three cases
are shown in \autoref{tab:fit_k_max}. Within error bars, one or two sigma,
the values
for $k_c$ agree with the above results which we obtained
from the data collapse.

\begin{table}
  \caption{\label{tab:fit_k_max} Parameters $k_c$, $\beta_1$ and $\beta_2$
    obtained from the fit of the position of
    maximum variance $k_\mathrm{max}(N)$ according to \autoref{eq:fit_k_max}. }
  \begin{tabular}{c c c c c}
      {$\eta$} & {$z$} & {$k_c$} & {$\beta_1$} & {$\beta_2$} \\
      \hline
      2 & 2 & 2.38566(3) & -0.07(1) & 137(21) \\
      4 & 2 & 0.88487(2) & 0.076(6) & -68(5) \\
      4 & 4 & 2.31068(4) & -0.04(1) & 404(118) \\
  \end{tabular}
\end{table}

To summarise, our results speak in favour of a phase transition
from a unsaturable to a saturable phase at a critical average degree
$k_c$ which depends on the graph structure. The scaling of the phase
transition seems to be coverned by a universal exponent $\nu\approx 2$,
which is very different from the usual mean-field exponent $\nu=1/2$.

 \subsection{Algorithm running time}

\begin{figure}
  \includegraphics{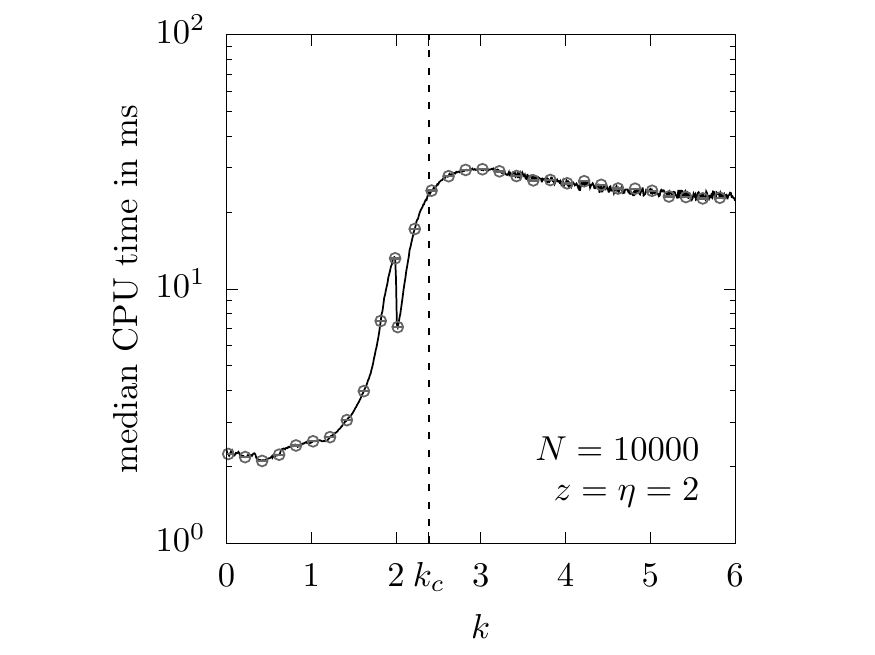}%
  \caption{\label{fig:Poisson_time_over_k} The median of the CPU
    time as function of user degree $k$, each value averaged over 1500 runs.
    The error bars are obtained using bootstrap resampling \cite{efron1994}.
    The circles show only some data points for better visualisation, 
but the black line connects all measured points. Note the bump at $k=2$
which is probably related to a fast heuristics used by the algorithm.
The vertical dashed line marks the critical point $k_c$ as obtained above.}
\end{figure}

We next analyse the running time of the matching algorithm when
varying the average user degrees, to see whether the phase
transition is reflected for this quantity. Unfortunately, the
package we used, the fastest open-source
matching algorithm implementation to our knowledge,
does not provide a machine-independent measure of the running
time. So he had to limit our self to
measure the CPU time. For this purpose, we used always the same
machine under the same conditions.
\autoref{fig:Poisson_time_over_k} shows the median of the  CPU
time for Erd\H{o}s-R\'{e}nyi graphs as the average degree increases.
Interestingly, the CPU time increases rapidly around the critical point,
a typical behaviour observed so far for a phase transitions in NP-hard
optimisation problems.
Note that
there appears to be a small non-monotonicity at $k=2$.
At $k=2$ each user has on average two servers
available and the number of users with this degree is maximal as compared
to all other degree values.
A possible explanation for this behaviour is that the matching algorithm,
which contains many sophisticated heuristics to speed up the
exact calculation, can improve the solutions here
substantially with just performing local changes for nodes of degree
exactly two.
For much larger values $k$, the median CPU-time decreases slowly,
even though there are more edges to handle. This is the case,
because the $z$-matchings become more and more degenerate in this region,
i.e., the algorithm has more feasible options to choose among.

 \subsection{Approximation algorithm}

\begin{figure}[ht]
  \includegraphics{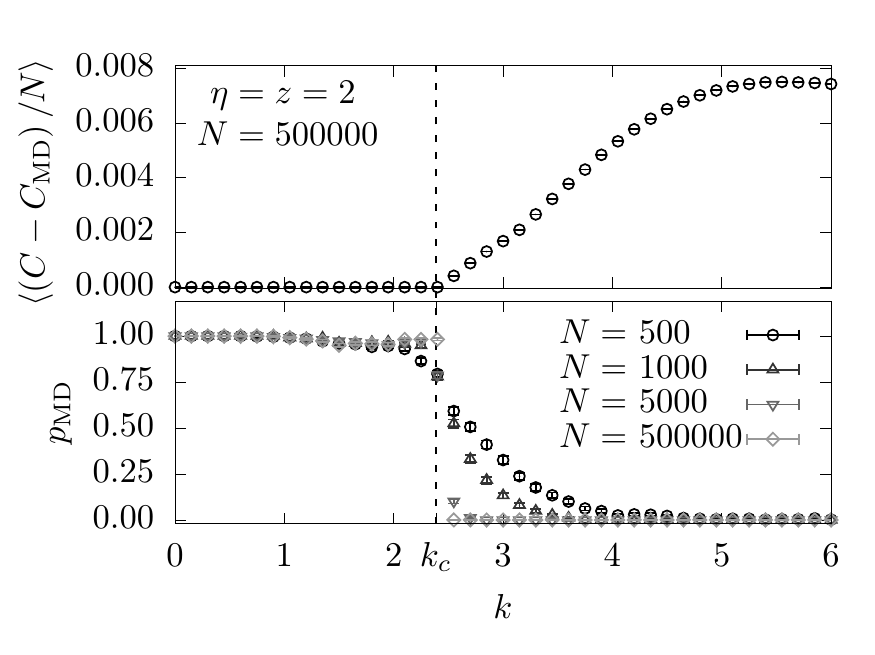}%
  \caption{\label{fig:Poisson_min_degree} Top: the difference of the
    capacity densities $C/N$ obtained from the exact matching algorithm and
    from the minimum degree (MD) heuristics on the same graphs.
    Bottom: $p_\mathrm{MD}$ as a function of $k$ for several system
    sizes $N$. The vertical dashed line marks the critical point $k_c$.}
\end{figure}

 The result of the running time of the exact algorithm shows that for
 small values of $k$, matchings can be obtained quickly. This could mean
 that in that region they are so simple to obtain such that
 even a non-exact but even faster algorithm is feasible.
Therefore, we compare the exact matching algorithm with a matching heuristics.
We considered the commonly used so called minimum-degree
heuristics \cite{Duff_2012}. It finds a matching by connecting
the nodes with lowest degree first, until no more nodes can be matched.
The basic idea is that for nodes with few neighbours, one has to find
a matching partner first, while nodes with many neighbours will still
find a partner even if many of their neighbours have been matched already.
Note that the heuristics has a linear running time.
In this context, we study $p_\mathrm{MD}$, i.e.,
the empirical estimated probability
that the minimum-degree heuristics obtains the same capacity as the
exact matching algorithm.
The top of \autoref{fig:Poisson_min_degree} shows how far
the capacity density $C/N$ obtained from the minimum degree heuristic
differs from the the one calculated by the exact matching algorithm.
On the bottom,
$p_\mathrm{MD}(k)$ is shown.
Both results are obtained over 500 different realisations for each
value of $k$.
Note that the average difference of the capacities between
both algorithms starts to grow for
$k \geq k_c$. In particular, $p_\mathrm{MD}$ undergoes for large systems an
almost step-wise transition from 1 to 0 near the critical point.
This means that for $k < k_c$ the heuristic finds solutions with are
well comparable to the exact solution. But for $k > k_c$ this is
no longer the case. Hence, the saturable-unsaturable phase transition
coincides with a kind of easy-hard transition with respect to a fast
heuristics, although the $z$-matching problem
is polynomially solvable everywhere.

\section{Summary and outlook}

We have studied the saturable-unsaturable phase transition
for the $z$-matching problem on bipartite Erd\H{o}s-R\'enyi random
graphs. Since the problem can be solved with exact algorithms
in polynomial time, we could study very large systems with good
accuracy, leading to high-precision estimates of the critical
points $k_c$ and of the critical exponent $\nu$ of the correlation
length.

We have also studied the running time of the exact algorithm and
found that the phase transition point is very close to the largest
change in the running time. Also, for the minimum-degree heuristics,
when studied for increasing node degrees, we find that the degree
beyond which the heuristics start to fail, seems to agree with
the critical point $k_c$. Thus, the saturable-unsaturable transitions
seems to coincide with strong changes of the algorithmic behaviour.
This was previously observed mainly for NP-hard optimisation or
constraints-satisfaction problems, not for polynomial problems.

Thus, for future work, it could be very interesting to study
other polynomial optimisation problems in a similar way and verify whether
phase transitions in connection with changes of the run time
are present. This could also apply to the investigation of other
ensembles of the $z$-matching problems, or other variants of matching.
This could lead to better understand the relation
between phase transitions and computational hardness
of optimisation or decision problems.

Finally, to understand this relation even better,
one could analyse the solution structure
for the $z$-matching problem. For this purpose one could
 extend the algorithm to allow for sampling of degenerate
solutions, possibly by introducing slight random changes to the
initially uniform edge weights, in order to make
so-far found solutions unfavourable. This would allow
to investigate the cluster structure \cite{jain1988} of the solution space.
Possibly one could in this way
observe changes of the solution-space structure in connection
with the saturable-unsaturable transition, as it has been the
case for NP-hard problems \cite{vccluster2004,krzakala2007,sat_cluster2010}.

\begin{acknowledgements}
  The simulations were performed at the 
  the HPC cluster CARL, located at the University of Oldenburg
  (Germany) and
    funded by the DFG through its Major Research Instrumentation Program
    (INST 184/157-1 FUGG) and the Ministry of
    Science and Culture (MWK) of the
    Lower Saxony State. 
\end{acknowledgements}

\bibliography{references}

\end{document}